\documentstyle[12pt,version2,aps]{revtex}

\draft
\begin{document}

\begin{title}
Superconductivity in the two-band Hubbard model
in Infinite Dimensions
\end{title}

\author{Antoine Georges*, Gabriel Kotliar**, and Werner Krauth***}
\begin{instit}
\begin{center}
* Laboratoire de Physique Th\'{e}orique de l'Ecole Normale Sup\'{e}rieure$^1$\\
24, rue Lhomond 75231 Paris Cedex 05; France\\
e-mail: georges@physique.ens.fr\\
** Serin Physics Laboratory, Rutgers University\\
 Piscataway, NJ 08855-0849, USA\\
e-mail: kotliar@physics.rutgers.edu\\
***Laboratoire de Physique Statistique de l'Ecole Normale Sup\'{e}rieure$^2$\\
24, rue Lhomond 75231 Paris Cedex 05; France\\
e-mail: krauth@physique.ens.fr\\
\end{center}
\end{instit}

\begin{abstract}
\noindent
We study a two-band Hubbard model in the limit of infinite dimensions, using
a combination of analytical methods and Monte-Carlo techniques. The normal
state is found to display various metal to insulators transitions as a
function of doping and interaction strength. We derive self-consistent
equations for the local Green's functions in the presence of superconducting
long-range order, and extend previous algorithms to this case. We present
direct numerical evidence that in a specific range of parameter space, the
normal state is unstable against a superconducting state characterized by a
strongly frequency dependent order-parameter.

\medskip
\noindent
$^1$ Unit\'{e} propre du CNRS (UP 701) associ\'{e}e \`{a} l'ENS et \`{a}
l'Universit\'{e} Paris-Sud

\noindent
$^2$ Laboratoire associ\'{e} au CNRS (URA 1306) et aux
Universit\'{e}s Paris VI et Paris VII

\medskip
\noindent

\end{abstract}
\hspace{2cm}

\newpage
\section{Introduction}
The discovery of high temperature superconductivity in the copper
based transition metal oxides has revived the interest in
superconductivity of strongly correlated electron systems.
Mechanisms of high temperature superconductivity which are not
phonon-mediated have been sought for a long time, and their existence
has remained a controversial subject. This
old problem cannot be settled by perturbative methods
\cite{littlewood}.

The large-dimensionality limit of the strong correlation problem has received
much recent attention \cite{MV}, \cite{janis}, \cite{ogawa},
\cite{GK},
\cite{REV},
\cite{jarrell}
\cite{RZK}
\cite{GK1}
\cite{GK2}
especially in the context of the Hubbard model. This is a nontrivial
limit of the strong correlation problem which can be solved
non-perturbatively.

In this paper we define a model with {\it two degrees of
freedom per unit cell}
which also has a well defined limit when the coordination number gets large.
On the atomic scale the interactions in this model are purely
repulsive. We show that the normal phase is
stable against  phase separation and displays interesting
metal-insulator transitions as a function of doping, and of the
interaction strength.

For the first time we give in this a paper a derivation of the
self consistent equations for the local Green's functions
in the presence of off-diagonal superconducting
long range order that become exact in the limit of infinite dimensions.
We then solve these equations by extending the Hirsch-Fye
Quantum Monte Carlo algorithm \cite{Hirsch}.

We present  evidence that in a well defined region of parameters
of the two-band model the normal state is
unstable against a superconducting state, characterized by a
strongly frequency dependent superconducting order parameter.
We observe that superconductivity in this model
is promoted in
regions of parameters
where two different atomic  configurations are
close in energy, that is in the  mixed valence regime.

\section{The Model}

We consider the following hamiltonian:
\begin{equation}
{\cal H}\,=\,
- \sum_{i\in A,j\in B,\sigma} t_{ij} d^{+}_{i\sigma} p_{j\sigma} + h. c.
+ \epsilon_p \sum_{j\in B,\sigma} p^{+}_{j\sigma} p_{j\sigma}
+ \epsilon_d \sum_{i\in A,\sigma} d^{+}_{i\sigma} d_{i\sigma}
+ U_d \sum_{i\in A} n^d_{i\uparrow} n^d_{i\downarrow}
\label{H}
\end{equation}
$(d_{\sigma},p_{\sigma})$ represent two atomic orbitals
on different sublattices $(A,B)$ of a
bipartite lattice. The ('copper') orbital $d_{\sigma}$ is strongly
correlated, while the ('oxygen') orbital $p_{\sigma}$ is uncorrelated.
Each site has identical connectivity $z$, so that the model describes
a '$CuO$'-type system. A two-dimensional three-band model has been
proposed by Emery \cite{Vic} and by Varma {\it et al} \cite{vsa}
as a minimal model of the copper-oxide $CuO_2$ planes. The model we
consider is similar in spirit, while having a non-trivial
large $d$ limit (as opposed to other possible generalizations of
multiband models which have somewhat degenerate large $d$ limits
\cite{trieste}, \cite{Gros}). Multiband models have been
investigated intensively,
but no clear numerical evidence for superconductivity in two dimensions
has so far been obtained \cite{muramatsu}, \cite{scalettar}.
In infinite dimensions several circumstances facilitate the numerical
work: the model reduces to an effective single-site model, in
which the thermodynamic limit is taken exactly, without the need
for finite-size extrapolations, and the Monte Carlo algorithm is free
from minus-sign problems, allowing us to reach lower temperatures than
in previous works \cite{muramatsu}\cite{scalettar}.

In  the absence of correlations ($U_d=0$), diagonalization of ${\cal H}$
yields two bands (bonding and antibonding):
$E_{\bf k}^{\pm}=\{\epsilon_p+\epsilon_d \pm
\sqrt{(\epsilon_p-\epsilon_d)^2+
4\epsilon_{\bf k}^2}\}/2$, where $\epsilon_{\bf k}$ is the Fourier
transform of $t_{ij}$. These bands are separated by a gap
$\Delta_0\equiv \epsilon_p-\epsilon_d$.
(Having in mind the copper-oxides in the
hole representation, we choose throughout this paper
$\epsilon_d<\epsilon_p$). For a total hole density
$n\equiv \sum_{\sigma} (<n_{d\sigma}>+<n_{p\sigma}>)$
smaller than $2$,
the $U_d=0$ ground state is a partially filled copper band and an empty
oxygen band. For $2<n<4$, the $U_d=0$ ground-state is a full d-band and
a partially filled p-band. The $T=0$ chemical potential is discontinuous as a
function of density at $n=2$, with $\mu(n=2^-)=\epsilon_d$,
$\mu(n=2^+)=\epsilon_p$. Hence, for $U_d=0$, this model describes a metal for
all densities except $n=2$, where one has a band insulator. The copper and
oxygen density of states have simple expressions for $U_d=0$:

\begin{equation}
N_d(\epsilon)=\sqrt{{{\epsilon-\epsilon_p}\over{\epsilon-\epsilon_d}}}\,
N(\sqrt{( \epsilon-\epsilon_p)(\epsilon-\epsilon_d)})
\,\,\,\,\,
N_p(\epsilon)=\sqrt{{{\epsilon-\epsilon_d}\over{\epsilon-\epsilon_p}}}\,
N(\sqrt{( \epsilon-\epsilon_p)(\epsilon-\epsilon_d)})
\label{freedos}
\end{equation}

\noindent
where $N(\epsilon) \equiv \sum_{\bf k} \delta(\epsilon-\epsilon_{\bf k})$.

The limit of infinite connectivity
$z\rightarrow\infty$ requires
a scaling of the hybridization $t_{ij}$ as: $t_{ij}=t_{pd}/\sqrt{z}$, so
that the density of states
$N(\epsilon)$
has a proper limit \cite{MV}. In practice, one may consider the
d-dimensional hypercubic lattice $(z=2d)$, for which:
$N(\epsilon)=1/\sqrt{2\pi}t_{pd}\,e^{-\epsilon^2/2t_{pd}^2}$
as $d\rightarrow\infty$ or the Bethe lattice with connectivity
$z$ for which $N(\epsilon)=\sqrt{4-(\epsilon/t_{pd})^2} / 2\pi t_{pd}$
as $z\rightarrow\infty$. We shall establish
the general equations for an arbitrary $N(\epsilon)$ but,
for the sake of simplicity, will perform all numerical simulations for the
Bethe lattice.

It is by now well-established that, in the limit $z\rightarrow\infty$, the
problem reduces to a single-site 'impurity' model supplemented by a
self-consistency condition.
Standard methods \cite{trieste}  that will not be reviewed here
allow us to derive the corresponding equations for the one-particle
Green's functions. In this section, we assume no long-range order of
any kind (magnetic or superconducting) and display equations valid in the
paramagnetic normal state.

Since only the 'copper' sites are
correlated, all the local  quantities
can be derived from an impurity model for those sites only.
The effective action of the impurity model is given by:

\begin{equation}
{\cal S}\,=\,
U_d\,\int^{\beta}_{0} d\tau\, n_{d\uparrow}(\tau)
n_{d\downarrow}(\tau)
- \int^{\beta}_{0} d\tau \int^{\beta}_{0} d\tau' \sum_{\sigma}
d_{\sigma}(\tau)
D_0^{-1}(\tau-\tau')\, d^{+}_{\sigma}(\tau')
\label{S}
\end{equation}

\noindent
We denote by
$D(\tau-\tau')\equiv -<Td(\tau) d^{+}(\tau')>_{{\cal S}}$ the {\it interacting}
Green's function calculated  with this action
and by
$\Sigma_d$  the impurity self-energy (seen as a functional of
$D_0$),
$\Sigma_d(i\omega_n)\equiv D_0^{-1}(i\omega_n)-D^{-1}(i\omega_n)$.
The self-consistency equation for $D_0$ then reads,
($\omega_n=(2n+1)\pi/\beta$):

\begin{equation}
D(i\omega_n)=\,\zeta_p(i\omega_n)\,\int d\epsilon \,
{{N(\epsilon)}\over{\zeta_p(i\omega_n)\zeta_d(i\omega_n)-\epsilon^2}}
\label{sc1}
\end{equation}

\noindent
where we have set set $\zeta_p(i\omega_n)\equiv i\omega_n+\mu-\epsilon_p$,
$\zeta_d(i\omega_n)\equiv i\omega_n+\mu-\epsilon_d-\Sigma_d(i\omega_n)$.

Once Eqs.(\ref{sc1}) ,(\ref{S}) are solved for  $D_0$ , the impurity self
energy
evaluated at the self consistent value of $D_0$  gives the
d electron lattice self energy
In the $(d_{\bf k\sigma},p_{\bf k\sigma})$ basis, it reads in matrix form:
\begin{equation}
{{1}\over{\zeta_p\zeta_d-\epsilon_{\bf k}^2}}\,\,\,\left(
\begin{array}{ll}
\zeta_p\,\,&\,\,\epsilon_{\bf k}\\
\epsilon_{\bf k}\,\,&\,\,\zeta_d
\end{array}
\right)
\label{gf}
\end{equation}

\noindent
In the $z\rightarrow\infty$ limit, self-energies become purely site-diagonal,
so that $\Sigma_d$ depends only on frequency and the $\Sigma_{pd}$ component
is absent. The absence of the diagonal component $\Sigma_p$ in eq.(\ref{gf})
comes from the simplifying assumption of an uncorrelated $p$-orbital. From
eq.(\ref{gf}), one sees that the self-consistency equation simply means
that the impurity model Green's function must coincide with the
on-site d-orbital Green's function:
$D(i\omega_n)=\sum_{\bf k} D({\bf k},i\omega_n)$. Also, the p-orbital
on-site Green's function is simply given by:

\begin{equation}
P(i\omega_n)=\,\zeta_d(i\omega_n)\,\int d\epsilon \,
{{N(\epsilon)}\over{\zeta_p(i\omega_n)\zeta_d(i\omega_n)-\epsilon^2}}
\label{p}
\end{equation}

It is straightforward to extend these equations to a model involving
a Hubbard repulsion on oxygen orbitals also. One then has to solve
simultaneously two separate impurity models, one for each orbital, and both
$\Sigma_d$ and $\Sigma_p$ enter the self-consistency equations
(\ref{sc1},\ref{p}). It is also possible to include a direct
oxygen-oxygen hopping $t_{pp}$ (with a $1/z$ scaling
as $z\rightarrow\infty$). For the sake of simplicity,
these additional terms  will not be considered in this paper.

Throughout the rest of this paper, we shall work with the $z=\infty$ Bethe
lattice with the semi-circular d.o.s. given above. This is motivated by the
very simple form taken by the self-consistency equations in this case:

\begin{equation}
D_0^{-1}\,=\, i\omega_n+\mu-\epsilon_d-t_{pd}^2 P(i\omega_n)
\qquad
P^{-1}\,=\, i\omega_n+\mu-\epsilon_p-t_{pd}^2 D(i\omega_n)
\label{sc2}
\end{equation}

\noindent
A direct way to derive these equations for the Bethe lattice is to integrate
out fermionic degrees of freedom on all lattice sites except a single one.
Because of the $z=\infty$ limit, only one-particle processes (involving the
full Green's functions $D,P$) are generated in this partial summation,
yielding the effective action eq. (\ref{S}), with $D_0$ explicitly given in
terms of $D$ by eq.(\ref{sc2}).

In order to solve the coupled problem defined by eqs.(\ref{S}), (\ref{sc2}),
we use a numerical method that has been described in detail elsewhere
\cite{jarrell} \cite{RZK}\cite{GK1} \cite{GK2}. It is
based on an iterative procedure: for a given imaginary-time $D_0(\tau)$,
the impurity model (\ref{S}) is solved (given a discretization of the
interval $[0,\beta]$ in $L=\beta/\Delta\tau$ time slices) using the
Hirsch-Fye quantum Monte-Carlo  algorithm. This produces a discretized
$D(\tau)$ which, after Fourier transformation, is inserted into
eq.(\ref{sc2}) in order to produce a 'new' function $D_0$. This process is
iterated until convergence is reached.

\section{Metal-Insulator transitions}

\subsection{The metal-insulator transition at $n=1$}

For a density $n=1$ this model displays
a transition from a metallic to an insulating state following a
scenario outlined by Zaanen, Sawatzky and Allen \cite{Zaanen}.

For very large $U_d/\Delta_0$
and $t_{pd}<<\Delta_0=\epsilon_p-\epsilon_d$, the $n=1$
ground state has all copper sites nearly singly occupied and all
oxygen sites nearly empty.
Hybridization between the two orbitals costs an energy $\simeq \Delta_0$,
while the kinetic energy gain is only $\simeq t_{pd}^2/\Delta_0$. Hence, for
$t_{pd}<<\Delta_0$, we have a {\it charge-transfer insulator} at $n=1$, with
a jump in the chemical potential $\mu(n=1^{+})-\mu(n=1^{-})\neq 0$.
The numerical  data for $n_d, n_p$ and the total density $n$
as a function of the chemical potential $\tilde{\mu} \equiv  \mu - \epsilon_d$
confirms these expectations.
In fig. \ref{fig1}, we show data at $\beta = 30$,
$U_d=8$, and $\Delta_0 = 4$.

In the
opposite limit ($t_{pd}>>\Delta_0$) the delocalization energy wins, and we
have a strongly correlated metal with a strong hybridization of 'copper' and
'oxygen' orbitals. A metal to charge-transfer insulator transition separates
these two regimes, at $(\Delta_0/t_{pd})_c=O(1)$ for large $U_d$.

In the opposite regime of weak correlations ($U_d<<\Delta_0$), the
metal-insulator transition at $n=1$ has the character of a Mott transition
within the copper band. Assuming that $t_{pd}<<\Delta_0$, most of the
$d$-orbital density of states is, for $U_d=0$, concentrated around
$\epsilon_d$, with a small bandwith of order $t_{pd}^2/\Delta_0$ (see
eq.\ref{freedos}) and the hybridization with the oxygen orbital is weak.
As $U_d$ is
increased, the lower copper band is gradually split by the interactions, and
a Mott transition occurs when $U_d$ becomes comparable to the bandwith
$t_{pd}^2/\Delta_0$, i.e. $(\Delta_0/t_{pd})_c\simeq t_{pd}/U_d$
for small $U_d$.
In fig. \ref{fig2} we show data at $\beta = 30$,
$U_d=1.5$, and $\Delta_0 = 4$ displaying this behavior.

Based on these arguments, the phase boundary separating the metallic and
insulating regimes at $n=1$ is expected to have the schematic shape
 \cite{Zaanen} described in fig. \ref{fig3}.

When $U \sim \Delta_0$, there is a crossover regime which to our knowledge
has not been investigated previously. Data in this region is shown in
fig. \ref{fig4}.
The  detailed investigation of these phase transitions is left for
future work.

\subsection{The metal to band insulator transition at $n=2$}

Interactions can also induce a transition from insulating
to metallic behavior at
a density $n=2$. At $U_d=0$ the system is a band insulator
with a band gap  $\Delta_0 \equiv \epsilon_p - \epsilon_d$.
Turning on the interactions has the effect of reducing the band gap.
Whenever the gap is less than the temperature, we expect metallic
behavior.

This change from insulating to metallic  behavior at $n=2$ is
clear in the $n$ vs  $\mu$ curves in figs \ref{fig1} \ref{fig2} and
\ref{fig4}.
In fig.\ref{fig2} the case  $U_d<\Delta_0$ is realized, for
which the atomic ground-state \cite{atomic} at $n=2$ has nearly all
copper sites doubly occupied and all oxygen sites empty.
We see gaps at $n=1$ and $n=2$ with
$\mu(n=1^{+})-\mu(n=1^{-})\simeq U_d$,
$\mu(n=2^{+})-\mu(n=2^{-})\simeq \Delta_0 - U_d$.
In fig \ref{fig1} the case  $U_d>\Delta_0$ is realized and the
corresponding atomic ground-state at $n=2$ \cite{atomic} has all copper and
oxygen sites singly occupied. Here,
gaps exist at $n=1$ and $n=3$, with
$\mu(n=1^{+})-\mu(n=1^{-})\simeq \Delta_0$,
$\mu(n=3^{+})-\mu(n=3^{-})\simeq U_d-\Delta_0$.

Hence an insulator to metal transition can be thought to occur
at $n=2$ at a critical
value of the ratio $U_d/\Delta_0 \sim 1$, which can be interpreted
as the upper Hubbard band hitting the $p$ level. In fig. (\ref{fig4}) such an
intermediate situation is realized.
Below, we find that very interesting physics arises in the
mixed valence regime
($U_d\simeq \Delta_0$), where
the strongly correlated copper level lies close to the uncorrelated
oxygen level.

Additional insights on this transition can be gained by looking at the
self-consistency equation (\ref{sc2}) for the special point $\mu=\epsilon_p$,
which
reads for low-frequency
($i\omega_n\rightarrow \omega+i0^{+}, \omega\rightarrow 0$):
\begin{equation}
D(0+i0^+)^{-1} + \Sigma_d(0+i0^+)=\epsilon_p-\epsilon_d + D(0+i0^+)^{-1}
\label{selfspecial}
\end{equation}
The low-frequency behaviour of $D$ and $\Sigma_d$ is severely constrained by
this equation,
which  allows two different solutions: i) $D(i0^{+})=0$
and ii) $\Sigma_d(i0^{+}) +\epsilon_d = \epsilon_p$. Case i) is
obviously realized for small $U_d$, while case ii) can be interpreted as
the effective $d$ level $\Sigma(i0^{+}) +\epsilon_d$ becoming
degenerate with the $p$ level.
Although we cannot obtain the numerical solution  at zero temperature we
find numerical evidence that one goes from i) to ii) as $U_d$ is increased.
This is readily seen in the data of fig.\ref{fig5} - \ref{fig8}
for four different values of the interaction strength. From these curves, it
is apparent that $\Sigma_d(i0^+)$ remains close to $\epsilon_p-\epsilon_d$ for
all values of $U_d$ larger than a critical value.
These figures display in fact the results of two subsequent  iterations of the
self consistency loop, which we display in order to highlight the
remarkable  accuracy of the Monte Carlo algorithm.

The qualitative behavior of $n$ vs $\mu$ can also be understood from
a Hubbard III-type solution of the coupled equations
(\ref{sc2}),
which takes into account the physics of high-energy processes in a
qualitatively correct manner \cite{RZK}, \cite{GK2}.
The appropriate generalization of the Hubbard III approach to the
two-band model inserts the following expression for $D(i\omega_n)$
\begin{equation}
D(i\omega_n)\,\simeq\,{{1}\over{2}}\{D_0\,+\,{{1}\over{D_0^{-1}-U_d}}\}
\label{appxS}
\end{equation}
into the self-consistency equation (\ref{sc2}),
to obtain the following closed equation for $D(i\omega_n)$:

\begin{equation}
\begin{array}{cc}
t_{pd}^4(x_d^2-U_d^2/4)\,D^3
+ t_{pd}^2(x_pU_d^2/2-2x_px_d^2+t_{pd}^2x_d)\,D^2 +\\
+ x^2-U_d^2x_p^2/4+t_{pd}^2x_px_d+t_{pd}^2x)\,D
- xx_p\,=\,0
\end{array}
\label{HIII}
\end{equation}

\noindent
where $x_d\equiv i\omega_n-\epsilon_d-U_d/2$,
$x_p\equiv i\omega_n-\epsilon_p$, $x\equiv x_px_d-t_{pd}^2$.

This simple approximation gives a rather satisfactory account of the
various metal-insulator transitions described above: zero-temperature
spectral densities are displayed in fig \ref{fig9} for various values of the
parameters.
It is unable however to account for the
low-energy quasiparticle excitations in the metallic regimes, since it just
ignores the Kondo effect in the impurity model eq. (\ref{S}).

\section{Superconducting properties}

\subsection{Dynamical pairing in infinite dimensions}

In this section we extend the $d=\infty$ formalism to incorporate
the possibility of superconducting long-range order. The equations that we
will establish here allow the investigation of the model within the
broken symmetry phase ( {\it cf} \cite{trieste} \cite{GK2}
where the Hubbard model was studied
within magnetically ordered states).
Let us define 'anomalous' Green's functions:
\begin{equation}
F_d(\tau-\tau')\equiv -T<d_{i,\downarrow}(\tau)d_{i,\uparrow}(\tau')>\,\,
F_p(\tau-\tau')\equiv -T<p_{i,\downarrow}(\tau)p_{i,\uparrow}(\tau')>
\label{Fdef}
\end{equation}
describing the formation of on-site pairs.
Singlet pairing corresponds to $F$ even: $F(\tau)=F(-\tau)=-F(\beta-\tau)$,
while $S_z=0$ triplet pairing corresponds to $F$ odd:
$F(\tau)=-F(-\tau)=F(\beta-\tau)$.
Allowing for a non-trivial time-dependence of $F$ is crucial. The underlying
physical idea is that on-site {\it equal-time} pairing is likely to be
strongly suppressed in the presence of a strong on-site repulsion but that
pairing involving a 'time-lag' between the paired holes may occur. This
idea dates back to Berezinskii's proposal \cite{berezinskii}
for triplet pairing in $^3He$, a generalization of which
has been recently considered for cuprate superconductors by Balatsky and
Abrahams \cite{balatsky} in the singlet case.

In the presence  of a non-zero $F$ it is convenient to work with Nambu spinors
$\Psi_d^{+} \equiv (d_{\uparrow}^{+}, d_{\downarrow})$ (similarly $\Psi_p$)
and with the matrix formulation of one-particle Green's functions:

\begin{equation}
{\bf D}(\tau-\tau') \equiv  -T<\Psi_d(\tau) \Psi_d^{+}(\tau')>\,=
\left(
\begin{array}{ll}
G_d(\tau-\tau')\,&F_d(\tau-\tau')\,\\
F_d(\tau-\tau')\,&-G_d(\tau'-\tau)\,\\
\end{array}
\right)
\label{gfsup}
\end{equation}

\noindent
With these notations, the kinetic term of the hamiltonian ${\cal H}$
reads:
$-t_{ij}\Psi_{d,i}^{+}\sigma_3 \Psi_{p,j}$ where $\sigma_3$ denotes the Pauli
matrix. Following the usual method, we integrate out fermionic variables
on all sites except on a single copper site. The 'impurity' action
obtained in this way now reads:

\begin{equation}
{\cal S}_{sup}\,=\,U_d\,\int^{\beta}_{0} d\tau\, n_{d\uparrow}(\tau)
n_{d\downarrow}(\tau)
- \int^{\beta}_{0} d\tau \int^{\beta}_{0} d\tau'
\Psi_d^{+}(\tau) {\bf D_0}^{-1}(\tau-\tau') \Psi_d(\tau')
\label{Ssup}
\end{equation}
where ${\bf D_0}$ is given in terms of ${\bf D}$ and ${\bf P}$
by the self-consistency
equations:

\begin{equation}
\begin{array}{ll}
{\bf D_0}^{-1}(i\omega_n) = i\omega_n + (\mu-\epsilon_d)\sigma_3
-t_{pd}^2 \,\sigma_3 {\bf P} (i\omega_n) \sigma_3\,\\
{\bf P}^{-1}(i\omega_n) = i\omega_n + (\mu-\epsilon_p)\sigma_3
-t_{pd}^2 \,\sigma_3 {\bf D} (i\omega_n) \sigma_3
\end{array}
\label{scs}
\end{equation}
We can account for an externally applied dynamic
pairing field $\Delta_d(i \omega_n)$ on all copper sites
in the original lattice problem by adding a forcing term
\begin{equation}
\left(
\begin{array}{ll}
0\,\,&\,\,\Delta_d(i \omega_n)\\
\Delta_d(i \omega_n)\,\,&\,\,0
\end{array}
\right)
\label{forcing}
\end{equation}
to the r. h. s. of eq. (\ref{scs}).

The impurity action eq. (\ref{Ssup})
describes an Anderson impurity in a superconducting medium.
Since this problem, even with static pairing
\cite{muller} turns out to be highly non trivial,
we can expect that the self consistent solution of
eq. (\ref{scs}) will allow very intricate densities of states.

\subsection{Algorithm in the presence of pairing\cite{change}}

In the present paragraph we show how the  Hirsch-Fye algorithm \cite{Hirsch}
can be generalized in the presence of off-diagonal
terms (in the spin indices) in the Green's function eq. (\ref{gfsup})
which, after the usual Trotter breakup, becomes a $2L \times 2L$ matrix.


The quartic term in the action is decoupled, as usual,
by the introduction of Ising variables.

\begin{equation}
\exp[-\Delta \tau U n^\uparrow n^\downarrow -
( n^\uparrow + n^\downarrow)/2=
Tr_\sigma \exp [\lambda \sigma(l)[n^\uparrow - n^\downarrow]]
\label{decoup}
\end{equation}

According to the prescription given by Soper \cite{Soper}, all terms in
the action have to be normal-ordered (here, in $\Psi^{+},\Psi$ with
$\Psi^{+}(\tau)= (\Psi^{+}_1(\tau),\Psi^{+}_2(\tau))
= (c_{\uparrow}^{+}, c_{\downarrow}$)),
before performing the discretization. Consequently, eq. (\ref{decoup}) is
written as
\begin{equation}
Tr_{\sigma} \exp[\lambda \sigma[\Psi_1^{+}
\Psi_1 + \Psi_2^{+} \Psi_2] \exp[-\lambda \sigma(l)]
\end{equation}

The Green's functions for different spin configurations are related by
the following Dyson equation:
\begin{equation}
{\bf D}_{\sigma'} = {\bf D}_{\sigma} +
( {\bf D}_{\sigma} - 1)(\exp(W' - W) -1)  {\bf D}_{\sigma'}
\label{Dyson}
\end{equation}
where ${\bf D_{\sigma}(\tau,\tau')}$ is the full $2L \times 2L$ matrix Green's
function of $\Psi$ for a given spin configuration $\sigma(l)$, and
$W_k = W_{k+L} \equiv \lambda \sigma(k)$.
Eq. (\ref{Dyson}) is derived in the same way as the one in ref. \cite{Hirsch},
and differs from that expression only in a sign for the down-spin sector.
The Green's function of the model is given by
\begin{equation}
{\bf D}(\tau,\tau')=\frac{\sum_{\sigma}
{\bf D}_{\sigma}(\tau,\tau')Det({\bf D}_\sigma^{-1})
\exp(-\lambda\sum_{l}\sigma(l))}{\sum_{\sigma} Det({\bf D}_\sigma^{-1})
\exp(-\lambda\sum_{l} \sigma(l))}
\end{equation}
The statistical weight of a configuration is thus given by the product of
the usual fermion determinant, and of a scalar factor, which arises from
the commutator in eq. (\ref{decoup}).

As in the Hirsch-Fye algorithm
\cite{Hirsch},
we can use the Dyson equation to calculate the ratio $R$ of statistical weights
for flipping a spin $\sigma(k) \rightarrow \sigma'(k)= -\sigma(k)$,
\begin{equation}
 R=R_{\uparrow} R_{\downarrow}  \exp(-2 \lambda \sigma[k])
\label{detrat}
\end{equation}
\begin{equation}
 R_{\uparrow}= 1+[1-{\bf D}(k,k)] f
\end{equation}
\begin{equation}
 R_{\downarrow}=1+[1- {\bf D}(k+L,k+L) -{\bf D}(k+L,k) {\bf D}(k,k+L) f /
(1 +(1-{\bf D}(k,k)) f)] f
\end{equation}
with $f= [\exp(-2 \lambda \sigma(k)) -1]$.
Once
a flip of spin $\sigma(k)$ is accepted, all the elements
of the $2L \times 2L$ Green's function are
updated in a 2-step procedure, corresponding to the flip
of  $W_k$ and 	 $W_{k+L}$.
\begin{equation}
{\bf D}'(l,l')={\bf D}(l,l') + ({\bf D}(l,k)-\delta(l,k)) f
/ (1+(1-{\bf D}(k,k))f) {\bf D}(k,l')
\label{upd1}
\end{equation}
\begin{equation}
{\bf D}''(l,l')={\bf D}'(l,l') + {\bf D}'(l,k+L)-\delta(l,k+L)) f/
(1+(1-{\bf D}'(k+L,k+L)) f) {\bf D}'(k+L,l')
\label{upd2}
\end{equation}

We have extensively compared the algorithm with exact diagonalization
results for an impurity interacting with a few conduction electrons in
the presence of explicit pairing terms $d^{+}_{\sigma}d^{+}_{-\sigma}$,
and conclude that the algorithm is correct, and practical.
In fig. \ref{fig10}
we show data of one such realization of the superconducting Anderson
model for $\Delta \tau= 1, .5, .25, .125$, and $0$ (exact diagonalization).
Notice the quadratic convergence.

\subsection{Numerical Results}

To investigate the
transition into the
superconducting state at finite temperature
we define a pairing susceptibility
$\chi_p(i\omega_n, i\omega_m)$
\begin{equation}
F_d(i\omega_n)=
\sum_{m} \chi_p(i\omega_n, i\omega_m)
\Delta_d(i\omega_m)\quad
\Delta_d(i\omega_m) \rightarrow 0
\label{chip}
\end{equation}

This susceptibility is a
well defined symmetric matrix
and is finite in the normal state. The divergence of its
largest eigenvalue signals the transition into the superconducting state.
In our approach it is natural to express $\chi_p$ in terms
of two operators, $\lambda_K$ and $\alpha$.
$\lambda_K(i\omega_n, i\omega_m) $  describes
the pairing response of the impurity model eq. (\ref{Ssup})
\begin{equation}
F_d(i\omega_n)=
\sum_{m} \lambda_K(i\omega_n, i\omega_m)
 [{G_{0d}}^{-1}]_{12}(i\omega_m) \quad
 [{G_{0d}}^{-1}]_{12}
 \rightarrow 0
\label{lambda}
\end{equation}
while the diagonal
operator $\alpha$ describes the response of the self consistent
medium to the impurity site and can be obtained from eq.(\ref{scs}) as
$\alpha(n,m)=\delta_{n,m} t_{pd}^4 {P^*}(i\omega_n)P(i\omega_n)$.

Clearly, $\lambda_K$, $\alpha$ and $\chi_p$ are related
by the following equation
\begin{equation}
\chi_p=[I-\lambda_{K}  \alpha]^{-1} \lambda_K
\end {equation}
When the largest eigenvalue of $\lambda \equiv \lambda_{K} \alpha$
approaches one,
the susceptibility diverges.
It is clearly more accurate to study the behavior of the largest
eigenvalue of $\lambda \equiv \lambda_k \alpha$ and we do that by the power
method using the
full algorithm.
To do this, we run the
algorithm described
in the previous section
a large number of times, starting with a small value of the off-diagonal
term $F(\tau,\tau')$ of the Green's function. In the normal state,
$F(\tau,\tau')$ will  converge to zero and two subsequent
 runs will yield two functions
functions  $F(\tau,\tau')$and $F'(\tau,\tau')$  obeying
\begin{equation}
F'(\tau,\tau')= \int d\tau'' \lambda(\tau,\tau'')
F(\tau'',\tau') \sim
\lambda_{max}
F(\tau,\tau')
\end{equation}

$\lambda_{max}$ is
the largest eigenvalue of the
matrix $\lambda$ ({\it cf} eq. (\ref{lambda})  and
$F'(\tau-\tau')$ converges to the corresponding right
eigenvector, which is also the eigenvector corresponding
to the most unstable eigenvalue of the susceptibility
matrix.


In fig (\ref{fig11}), we show
$\lambda_{max}$ vs n at a relatively high temperature
$\beta=15$ in the
singlet sector, {\i.e.}, after explicitly anti-symmetrizing the
functions $F(\tau)$ and $F_0(\tau)$ with respect to $\beta/2$.
In order to  assess
 the importance of finite $\Delta \tau$ effects, we have plotted
$\lambda_{max}$ for $L=16,32,64$.
Let us stress that this extrapolation is the only error in our
calculation, since
the statistical uncertaintly and the uncertainty stemming from the
self-consistency condition are extremely small.
All of our data is consistent with quadratic convergence in
$\Delta \tau ^2$
for small values of $\Delta \tau$ which we have clearly
established for the Green's function itself.

The susceptibility
seems to have a local maximum close to the density $n=1$, and is again
large at around $n=2$.
We notice that the tendency towards superconductivity is largest
near half filling and near $n\sim 2$, {\it i.e.} in the mixed valence regime.

We have been able to produce  perfectly converged
superconducting solutions of the meanfield equations in the
singlet  case for temperatures within our reach.
The solution for $F(\tau)$, at temperatures
just below
$T_c(\Delta \tau) $
is very similar to the largest eigenvector
of  the susceptibility matrix above $T_c$.
The finite value of $\Delta \tau$ produces large finite size
effects in the value of
$T_c(\Delta \tau) $.
At this point we have established  that $\beta_c(L=64)=37$,
$\beta_c(L =128)=59$
and $\beta_c(L =192)=75$.
A reliable finite size scaling of this quantity to
determine the value of the critical temperature requires
larger simulations and  will be described
elsewhere.
Here we use the fact that the largest eigenvector of the susceptibility
matrix converges very well to discuss the nature
 of the possible superconductiviy in this system.
 Of particular importance is the different and strong
frequency dependence on the copper and the oxygen sites.

Examples of such numerical solutions
are given in fig \ref{fig12}
for $128$ time slices, at the value which seems to be most favorable
for the appearance of superconductivity $\Delta_0 = 4$.
The Fourier transform of the order parameter
on the copper site near $T_C$ are displayed
in fig \ref{fig13}. Notice (inset of fig \ref{fig12}) that the pairing
amplitude is much larger on oxygen sites than on copper sites, as expected,
and that the equal-time pairing $F(0)$ is nearly zero on copper sites.

\section{Conclusion}
In this paper we introduced a controlled approach to the study of
the occurrence of superconductivity in strongly correlated systems.
We extendended the mean field approach
to the strong correlation problem  of ref. \cite{GK}
to derive a self consistent
picture of a correlated
superconductor which becomes exact
in infinite dimensions.
As in the corresponding treatment of the normal state (\cite{GK})
an impurity model plays the role of the mean field
hamiltonian.
The
impurity model corresponding to
a correlated superconductor
is  an Anderson impurity model  embedded in a superconducting
medium. The quantities that correspond most
closely to the Weiss field is
now a set of two functions that describe
the diagonal and the off diagonal response of
the effective medium. A numerical solution of the
self consistent equations for the
Green's functions revealed a strong frequency
dependence of the superconducting order parameter.

Using this method we investigated  the most promising
regions for high temperature superconductivity in
the present version of the extended Hubbard model
in large dimensions. The largest tendencies
towards superconductivity were found  when the
normal phase is is in a mixed valence regime.

There are several natural extensions of this work. One can study
the effect of adding extra interactions in the hamiltonian on the
superconducting transition temperature of the system.
While the critical temperatures we found are relatively high,
a better understanding of the behaviour of the solution of our
equations may lead to higher transition temperatures.
Finally, as emphasized
 throughout  the text, larger simulations
are called for.
They should allow the finite size scaling analysis
which can fully elucidate the existence of the superconducting
state proposed here.
A completely open problem is the study of the physical properties of this
superconducting state. We stress that because of the strong
correlations a non perturbative treatment like the one outlined in
our paper is necessary for tackling this problem.

\section{Acknowledgments}
W. Krauth would like to thank the physics department at
Rutgers University for its kind hospitality.
G. Kotliar acknowledges the support of the NSF under
grant DMR 922-4000 and the hospitality of the theoretical
physics group at the ENS where this work began.

\newpage
{\bf Figure Captions}
\begin{enumerate}
\item{}
\label{fig1}
$n_p$ (full line) $n_d$ (dotted line) and  $n$ (dashed line) {\it vs}
$\mu-\epsilon_d$ at $\Delta_0 = 4$ $U_d=8$, and $\beta= 30$.
\item{}
\label{fig2}
$n_p$ (full line) $n_d$ (dotted line) and  $n$ (dashed line) {\it vs}
$\mu-\epsilon_d$ at $\Delta_0 = 4$ $U_d=1.5$, and $\beta= 30$.
\item{}
\label{fig3}
The Zaanen Sawatzky Allen phase  diagram . The special point occurs
when $ U_d \sim \Delta_0$
\item{}
\label{fig4}
$n_p$ (full line) $n_d$ (dotted line) and  $n$ (dashed line) {\it vs}
$\mu-\epsilon_d$ at $\Delta_0 = 4$ $U_d=4.5$, and $\beta= 30$.
\item{}
\label{fig5}
$Im(G(i\omega_n))$ (full line) and $Re(\Sigma_d(i\omega_n))$ (dotted line)
for $U_d=1.5$, $\mu-\epsilon_d=4$,
$\Delta_0 = 4$, and $\beta= 45$ ($L=128$).
\item{}
\label{fig6}
$Im(G(i\omega_n))$ (full line) and $Re(\Sigma_d(i\omega_n))$ (dotted line)
for $U_d=2.5$, $\mu-\epsilon_d=4$,
$\Delta_0 = 4$, and $\beta= 45$ ($L=128$).
\item{}
\label{fig7}
$Im(G(i\omega_n))$ (full line) and $Re(\Sigma_d(i\omega_n))$ (dotted line)
for $U_d=4.5$, $\mu-\epsilon_d=4$,
$\Delta_0 = 4$, and $\beta= 45$ ($L=128$).
\item{}
\label{fig8}
$Im(G(i\omega_n))$ (full line) and $Re(\Sigma_d(i\omega_n))$ (dotted line)
for $U_d=6.5$, $\mu-\epsilon_d=4$,
$\Delta_0 = 4$, and $\beta= 45$ ($L=128$).
\item{}
\label{fig9}
HubbardIII approximation for the one-particle density of states
$\rho_d(\omega)$ {\it vs} $\omega$ for $\Delta_0=4$, and
$U_d=.25, 3, 4.5, 8$
\item{}
\label{fig10}
Green's function $G(\tau=0)$and $F(\tau=0)$ in one
realization of the superconducting Anderson
model for $\Delta \tau= 1, .5, .25, .125$, and $0$ (exact diagonalization).
We have  observed quadratic convergence of the Green's functions for any value
of $\tau$ and a wide range of parameters.
\item{}
\label{fig11}
$\lambda$ {\it vs} $n$ at $\beta= 15$ and
$L=16$ (upper curve), $32$, and $64$ (lower curve) at the special point
$\Delta_0 = 4$,
$U_d=4.5$, $\mu-\epsilon_d=4$,
\item{}
\label{fig12}
Green's functions $G_d(\tau)$ and $F_d(\tau)$ at the special point
$\Delta_0 = 4$, $\mu-\epsilon_d=4$, at $U_d=4.5$.
$\beta = 64$ and $L=128$. The inset shows
$F_d(\tau)$
and $F_p(\tau)$.
\item{}
\label{fig13}
The real part of $F_d(i\omega_n)$vs $\omega_n$
on the special point at
$L=128$ and inverse temperatures
$\beta=64$(full line),
$\beta=70$(dotted  line),
and $\beta=80$(dashed  line),
\item{}
\label{fig14}
Imaginary part of $G_d(i\omega_n)$
on the special point at
$L=128$ and inverse temperatures
$\beta=64$(full line),
$\beta=70$(dotted  line),
and $\beta=80$(dashed  line).

\end{enumerate}

\begin{references}
\bibitem{atomic} Note that, in the atomic limit $t_{pd}=0$, the total
hole density $vs.$ chemical potential is given by:
$n_{atomic}=\Theta(\mu-\epsilon_d)+\Theta(\mu-(\epsilon_d+U_d))+
2\Theta(\mu-\epsilon_p)$.
\bibitem{change} In this section we follow the Hirsch and Fye notation
and define the Green's functions with an extra minus sign relative to
the other sections.
\end{references}
\end{document}